# A Nano-satellite Mission to Study Charged Particle Precipitation from the Van Allen Radiation Belts caused due to Seismo-Electromagnetic Emissions

Nithin SIVADAS[1], Akshay GULATI[1], Deepti KANNAPAN[1], Ananth Saran YALAMARTHY[1], Ankit DHIMAN[1], Arjun BHAGOJI[1], Athreya SHANKAR[1], Nitin PRASAD[1], Harishankar RAMACHANDRAN[1], R. David KOILPILLAI[1]

[1]*Indian Institute of Technology Madras, India;*
iitmsat@iitm.ac.in

In the past decade, several attempts have been made to study the effects of seismo-electromagnetic emissions - an earthquake precursor, on the ionosphere and the radiation belts. The IIT Madras nano-satellite (IITMSAT) mission is designed to make sensitive measurements of charged particle fluxes in a Low Earth Orbit to study the nature of charged particle precipitation from the Van Allen radiation belts caused due to such emissions. With the Space-based Proton Electron Energy Detector on-board a single nano-satellite, the mission will attempt to gather statistically significant data to verify possible correlations with seismo-electromagnetic emissions before major earthquakes.

## 1. Introduction

Indian Institute of Technology Madras (IITM) is developing a nano-satellite in technical collaboration with the Indian Space Research Organization (ISRO) to measure the fluctuations in charged particle fluxes precipitated from the radiation belts to Low Earth Orbit (LEO). These fluctuations are mainly caused due to leakage or precipitation of the charged particles from the Van Allen radiation belts to LEOs [1].

There are several processes that cause precipitation of the trapped charged particles from the radiation belts. These processes are not clearly understood yet, but the primary cause is thought to be electromagnetic fluctuations in the radiation belt [1]. These electromagnetic fluctuations can be produced through solar-magnetic storms, lightning storms, man-made electromagnetic emissions and seismic activity [1].

For the past two decades, researchers have been studying electromagnetic precursors to earthquakes and its effect on the upper ionosphere and the radiation belts [2,3,4,5,6,7,8,9]. Studying these effects is one of the primary objectives of this mission. Understanding these phenomena may aid in the development of a global earthquake warning system in the future.

### 1.1. Phenomenon under study

Correlations between fluctuations in the charged particle fluxes in Low Earth Orbit and seismic activity have been reported in literature. The physical processes that link fluctuations in the charged particle flux and seismic activity are still open to debate, but possible physical models have to be tested in order to make any progress in this area of study.

Among the several hypotheses that can be found in literature, the following physical model of the phenomena suggested by Sgrigna [10] will be tested and evaluated by the IITM nano-satellite mission. During the development of an earthquake, Ultra Low Frequency (ULF) (f <5 Hz) / Extra Low Frequency (ELF) (f = 30 -300 Hz) electromagnetic waves have been observed on ground and Low Earth Orbit altitudes [11,12,13,14]. These seismo-electromagnetic emissions, produced by direct or indirect means, travel through the atmosphere and are captured near the ionosphere-magnetosphere transition region and continue their journey along the geomagnetic field lines as Alfven waves[1] (a type of magneto-hydrodynamic wave). These waves, near the inner Van Allen radiation belt boundary, resonantly interact with the trapped particles in the radiation belt and cause their precipitation. This precipitation may be observed a few hours before the manifestation of an earthquake by satellites in LEOs as particle bursts (sudden increase in the particle counting rates). This phenomenon is described in detail in Section 2.

### 1.2. Previous missions and studies

The first time a correlation between the precipitation of high-energy particles from the upper ionosphere and seismic activity was reported in the literature was in the late 1980s based on results obtained from the MARIA experiment on-board Salyut-7 (launched in 1982) [2,3]. The study of high-energy particle flux variations just below the Van Allen belts was continued by using data obtained from the MARIA-2 magnetic spectrometer on-board MIR and the ELECTRON instruments on-board the satellites INTERCOSMOS-BULGARIA-1300 and METEOR-3 [4,5]. In [6] and [7], the authors found that variations in the high energy particle fluxes were observed a few hours before the main shocks of powerful earthquakes.

Studies to find these correlations using the data from the PET instrument on-board the SAMPEX satellite [10] and the MEPED instrument on-board the NOAA satellites [8] have also been done. In 2006, the ARINA instrument was launched aboard the Resurs-DK1 satellite [9] with the aim of "studying a seismo-magnetospheric phenomenon of the generation of high-energy charged-particle fluxes in the near-Earth space a few hours before earthquakes". In 2012, NASA launched two probes called the Radiation Belt Storm Probes [15] (now called the Van Allen probes) which had a particle detector on-board called REPT whose energy range overlaps significantly with that of the PET probe on-board the older SAMPEX satellite. These probes discovered a short-lived third radiation belt beyond the second one. They provided a better mapping of the radiation belt regions than any before and the data from these can also be analyzed and compared with the SAMPEX data to check for correlations between seismic activity and variations in the flux of high-energy particles.

[10] contains a detailed study of data from SAMPEX that specified the particle energies where correlations can most likely to be found. The South Atlantic Anomaly and other non-seismic sources of particle precipitation were systematically excluded from the analysis. The authors specified the ideal characteristics that a mission and the associated detector should have to carry out a correlation study. They chose the SAMPEX PET as it best matched their requirements. The data was binned and the correlation between seismic activity and particle bursts was studied. The authors found statistically significant correlations approximately 4 hours before an earthquake.

In [8], the analysis method used in [10] was modified so that it could be used to analyze electron flux data from the NOAA satellites which have higher orbits and different energy ranges. Here, the authors found significant correlations about 4 days before an earthquake. They also found long term correlations of up to 150 days.

However, the authors of [8] acknowledge the analysis may not be entirely accurate and various factors may, in fact, give falsely positive correlations. Also, the MEPED detector on-board the NOAA satellites has a smaller aperture and energy range than the PET on SAMPEX. Thus, based on the requirements for a mission given in previous studies, the data from the MEPED detector may not be optimal for determining correlations.

Considering the drawbacks and limitations of previous missions and studies, we have attempted to optimize our mission parameters to best fit the requirements recommended by Sgrigna [10]. This will allow us to study the effect of seismo-electromagnetic emissions on charged particles better and look for correlations with a higher degree of accuracy.

### 1.3. Current mission's contribution

The payload of the IIT Madras nano-satellite is a high energy charged particle detector called Space-based Proton Electron Energy Detector (SPEED). SPEED, with its large active area (~ 625 $cm^2$), can record extremely sensitive charged particle flux measurements compared to any of the previous missions. The mission is also designed to point the detector axis along the magnetic field line, so as to measure charged particles that are precipitated from the radiation belt, and with a pitch angle (the angle between the particle's velocity vector and magnetic field vector) which is lower than the loss cone angle (critical pitch angle of the charged particle below which the particle is precipitated from the Van Allen radiation belts). Lastly, the energy range of SPEED (17 to 100 MeV protons and 1-15 MeV electrons) is higher than most of the previous missions. This energy range lies within the range of charged particle energies which can undergo bounce and cyclotron resonance[1] with the ULF/ELF waves generated by seismic activity.

Simulations of low frequency wave interactions with the trapped charged particles in the radiation belts are also being developed at IIT Madras to predict the nature of charged particle precipitation that may be observed at low earth orbit altitudes. The data from the IIT Madras nano-satellite mission can also be used to test the predictions of these models.

### 1.4. Outline of the paper

The rest of the paper is organized as follows. Section 2 contains a brief description of the phenomena we plan to study using the IITM nano-satellite mission. Section 3 describes the scientific objectives, mission requirements and implementation, method of data analysis, and possible controlled experiments that can be performed during the mission life. Section 4 is devoted to the design specifications and unique features of the payload on-board the IITM Nano-satellite. The paper concludes by discussing the possible relevance of the results of this experimental mission on aiding the development of an earthquake warning system in the future.

## 2. Description of Phenomenon under Study

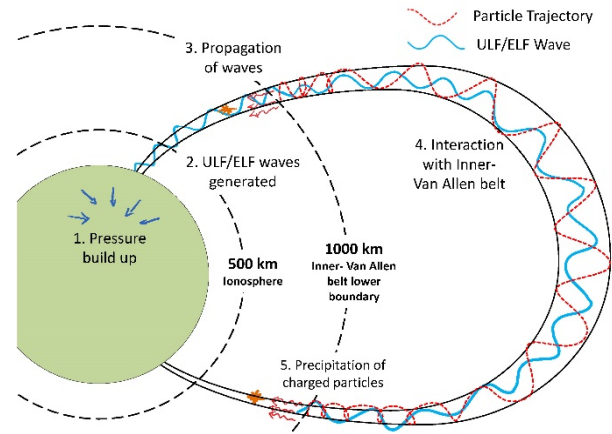

Figure 1: Summary of the phenomenon under study

The following paragraphs describe, in detail, the process by which particle bursts may be produced due to seismic activity.

### 2.1. ULF/ELF Wave Generation due to Seismic Activity at the Hypocenter

There are several theories that attempt to explain the production of low frequency electromagnetic waves before an earthquake. Among the mechanisms that are mentioned in the literature [11], the following are possible candidates for the generation of seismo-electromagnetic waves: Electro-kinetic mechanism [12], Surface dipole oscillations as the source of electromagnetic radiation [16], and Earth's crust acting as a charging electrical battery under increasing stress [17].

### 2.2. Propagation of ULF/ELF waves through the ground and the atmosphere

The electromagnetic waves generated by the seismic activity, travel from the hypocenter region and reach the ionosphere through the intervening ground and the atmosphere. During their propagation through the solid crust, the higher frequency contents of the SEME (Seismo-Electro-Magnetic Emissions) waves are severely attenuated. Thus, only the ULF/ELF waves are thought to reach the Earth's surface and propagate further into the near-Earth space, with least attenuation. [18]

### 2.3. Coupling OF ULF/ELF Waves with the Geomagnetic Field Line

Authors in [6,19,7,20] have proposed that such low frequency ULF/ELF waves (few mHz to a few hundred Hz), which are generated several hours before the main shock [21] are trapped in a channel (geomagnetic field tube) [22] made by the corresponding L-shell at an altitude of 300 - 500 km (this altitude corresponds to the maximum density region of ionosphere) [23]. From this region, these waves travel as Alfven waves further along the Earth's magnetic field line and reach the inner boundary of the inner Van-Allen radiation belt.

### 2.4. Interaction with Trapped Charged Particles in the Inner Van-Allen Belt

Wave-particle interactions in the inner magnetosphere take place if the resonance condition between the waves and the charged particles is satisfied [24]. ULF waves can undergo two types of resonance: bounce resonance and drift resonance. In bounce resonant interaction, the frequency of ULF waves coincide with the oscillation frequency of charged particles between the mirror points. While, in drift resonance, the frequency of ULF waves coincide with the frequency of revolutions of the charged particles around the earth due to drift.

### 2.5. Particle Precipitation

The resonant interactions of the ULF/ELF waves with the charged particles in the radiation belt change the particle's pitch angle and as a consequence, results in a decrease of *mirror point altitude* (the altitude at which the particle bounces back into the radiation belts) in comparison with stable trapped particles [21]. That is, *pitch angle diffusion* takes place and the particles precipitate from the inner Van-Allen radiation belt to the upper ionosphere. However, we assume that the particles remain in the same L-shell. These precipitated particles are observed as particle bursts by satellites.

### 2.6. Loss of the precipitated particles in the atmosphere

The precipitated particles drift longitudinally just like the trapped charged particles in the radiation belt. Lifetime of the longitudinal drift of these particles is determined by the particle loss rate during particle's interaction with residual atmosphere of the Earth. A life time of several tens of minutes is obtained for electrons and protons (of several tens of MeV) [1,21].

### 2.7. Locating the impending earthquake's epicenter

The coupling of the ULF waves with the inner magnetosphere is confined to a geomagnetic field tube of radial dimension 150 km (corresponding to a latitude variation of $1^0$ 30') [25]. The precipitated particles drift around the Earth confined within the same L-shell of its origin [10]. Thus a satellite orbiting the Earth can detect the particle bursts whenever it crosses the L-shell corresponding to the L-shell of the ULF wave coupling. The L-shell where the ULF/ELF waves get captured and the L-shell where the particle bursts occur should be more or less the same (according to Sgrigna [10], $\Delta L < 0.1$). Each L-shell corresponds to a pair of magnetic-shell iso-lines on the ground [9]. This means that, once the L coordinate of the particle burst is identified, the latitude of the future earthquake's epicenter can be determined. Furthermore, according to Galper [21], there is a distinction in the temporal profiles of counting rates for different longitudes of particle burst formation. This can be used to identify the longitude of the impending earthquake's epicenter.

## 3. Mission

### 3.1. Scientific Objectives

The scientific objectives of IITMSAT are as follows:
1. To investigate whether electromagnetic waves generated by seismic activity interact with the inner-Van Allen belt to precipitate detectable amounts of charged particles (protons and electrons)
2. To find out if particle bursts caused due to seismic activity can be distinguished from particle bursts due to other sources
3. To check for a correlation between charged particle bursts and seismic activity
   a. To study the relation between the time of occurrence of particle bursts and the time of occurrence of the earthquake main shock
   b. To investigate if location and time of observation of particle bursts can together be used to estimate the latitude and longitude of the earthquake's epicenter
   c. To investigate if the magnitude of the earthquake can be correlated to the size of the particle bursts
4. To characterize the background charged particle flux, and fluctuations in charged particle flux in the upper ionosphere

### 3.2. Mission Requirements

In order to satisfy the science objectives, the mission has to satisfy certain requirements. These mission requirements are summarized in Table 1.

*Science data products:* The mission is required to produce the following data for further analysis:
i. Energy spectrum of protons and electrons
ii. Pointing direction of the detector during measurement
iii. Position of the satellite during measurement to predict the location of the origin of seismic activity
iv. Time correlation of measurement

*Energy range and type of particle:* In the spectrum of electromagnetic waves emitted from the hypocenter of the seismic activity, ULF wave of frequencies 0.01 Hz to 10 Hz reach the magnetosphere with least attenuation [18]. This range of frequency approximately corresponds to the bounce frequency of protons of energy range 1 to 100 MeV and bounce frequency of electrons of energy range 1 to 15 MeV. Hence, the energy range of interest for protons is 1 to 100 MeV and that for electrons is 1 to 15 MeV.

*Energy, spatial and temporal resolution of the measurement:* The best signal to noise ratio can be achieved if temporal resolution is 0.4 seconds or lower, as this happens to be the duration of passage of the particle burst (of 50 MeV protons) through the satellite.

Seismic activity produces Alfven waves that travel along a geomagnetic field tube of diameter 300km [25] affecting the corresponding narrow range of L-shells (~0.046 L). Thus, knowing the location of particle bursts we can perform a latitude correlation of the epicenter. Hence the spatial resolution must be much less than 300km.

The energy information is required only to obtain and identify trends in the energy spectrum. Hence, an energy resolution of 5 MeV is assumed to be sufficient.

*Coverage:* The mission is aimed at monitoring charged particle precipitations under the inner Van-Allen radiation belts. This corresponds to an altitude of 600-800km above sea level. Further, at these altitudes, the range of the inner Van-Allen radiation belt varies from L-Shells 1.15 to 1.75. Therefore, the detector should carry out measurements in this desired range of L-Shells.

Table 1. Requirements on Mission

| Science data products | i. Energy spectrum of protons and electrons |
| --- | --- |
| | ii. Pointing direction of the detector |
| | iii. Position of the satellite |
| | iv. Time |
| Energy Range and Particle type | Protons: 1 to 100MeV |
| | Electrons: 1 to 15MeV |
| Resolutions | Energy < 5MeV |
| | Time < 0.4s |
| | Spatial < 15km |
| Coverage | L-shells: 1.15 L to 1.75 L |
| | Altitude: 600 to 800 km |
| Attitude | $0°$ with respect to the magnetic field line |
| Aperture | As large as possible within the constraints of the mission |
| Mission life time | 1 year |

*Attitude:* The particles with high pitch angles are those that are normally present in the upper ionosphere, and are not the ones precipitated from the radiation belts. Therefore, the best signal (particle with pitch angles close to $0^0$) to noise (particle with pitch angle close to $90^0$) ratio can be obtained if the detector is pointing along the magnetic field line. As a result, the detector opening face must point along the magnetic field line.

*Aperture:* The detector should have a large opening area to measure the fluctuations in the particle fluxes (or particle bursts) accurately. The area of the detector was fixed by assuming the temporal resolution (in this case being 0.1 seconds) and the maximum error that we can tolerate in identifying a particle burst from the background particle flux (in this case it is set to 0.1%).

*Mission lifetime:* According to [10], considerable statistical correlations have been observed with particle burst and earthquakes of magnitude M ≥ 5.0. As per the United States Geological Survey [26], on an average, 1469 earthquakes of M ≥ 5.0 may occur in one year. This is sufficient for the experiment to be conducted rigorously. Therefore, the mission life is designed to be a minimum of 1 year.

### 3.3. Implementation

In order to meet the mission requirements, the project must have at least one high energy particle detector capable of measuring the required energies on-board a space-craft skimming the L-shells just below the inner-Van Allen radiation belt [10]. This minimum requirement can be met by a single satellite, in the appropriate orbit around the Earth, which can transmit the acquired measurements to a ground station for further analysis.

Further, to meet the coverage requirements, the satellite has to pass through the desired range of L-shells (1.15L to 1.75L). This L-shell range of interest is contained within the latitudes 17°N to 40°N and 17°S to 40°S. Therefore, an inclination of 40° is preferred for maximum science data quality. Also, to improve the sensitivity of charged particle flux measurements the active opening area of the high energy particle detector was chosen to be as high as possible within the constraints of the nano-satellite structure.

### 3.4. Data analysis

In order to meet the mission objectives, the project must gather data from other external sources which help in the correlation of seismic activities with particle flux anomalies. Some of these include:
i. ULF/ELF measurements from satellites such as DEMETER, TwinSat
ii. ULF/ELF measurements from seismic activity monitoring stations on ground
iii. Solar activity measurements from satellites such as GOES
iv. Lightning storm activity data from weather monitoring systems
v. Data on the magnitude, location and time of earthquakes from seismic activity monitoring stations

Figure 2 qualitatively explains the procedure for analyzing the data. A similar procedure was adopted by Sgrigna [10] for analyzing the data from the SAMPEX-PET detector.

The particle burst data obtained from the mission is first filtered to exclude charged particles bursts collected from non-seismic sources. After that, the data is analyzed by using the proposed model of the phenomenon to identify the earthquake epicenters. Using these estimated locations of earthquake epicenters and the external data sources, a statistical analysis for the correlation of particle bursts with earthquakes can be carried out.

### 3.5. Possibility of controlled experiments

Coupling between Very Low Frequency (VLF) waves and the magnetosphere have been established by the experiments described in [27]. Here, man-made VLF waves were generated with a VLF transmitter and associated precipitation of radiation belt electrons (caused due to the coupling) was observed on-board a satellite. This provides direct evidence for the coupling of low frequency waves with the geomagnetic field lines and their penetration into the radiation belts. Such controlled experiments will provide more insight into wave-particle interaction phenomena within the magnetosphere.

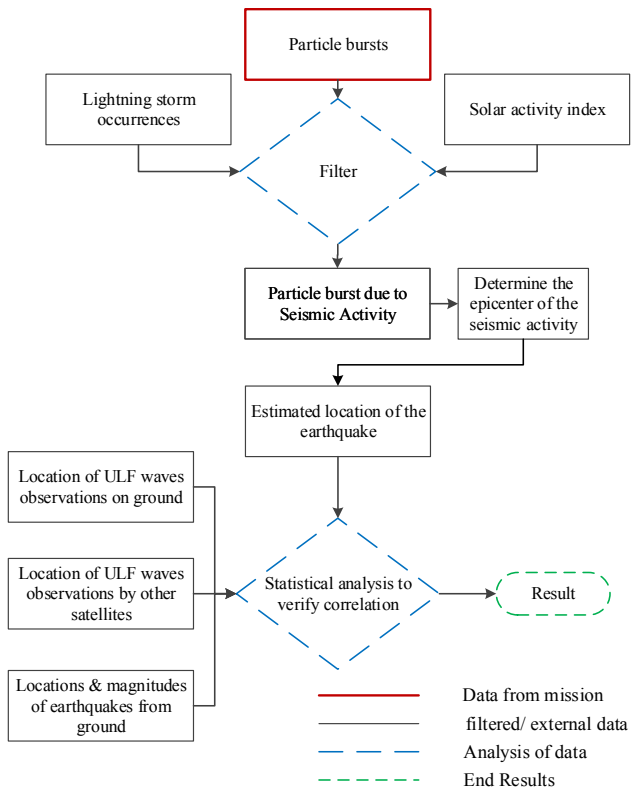

Figure 2: Procedure for analyzing the data obtained from the mission

## 4. Payload: Space-Based Proton Electron Energy Detector (SPEED)

### 4.1. Design description

SPEED is a plastic scintillator based charged-particle detector capable of measuring the energy spectrum of protons and electrons whose energies range between 17-100MeV and 1-15 MeV respectively. Charged particles lose energy in the plastic scintillator slabs, producing flashes of low amplitude light, which are termed as 'scintillations'.

The light from the scintillations is routed to the photomultiplier tubes (PMT's) located underneath the detector by directly coupling to the PMT or by using Wavelength Shifting Fibers (WLS). WLS fibers are used in the design to satisfy space constraints specific to the nano-satellite. Once the PMT's convert the light into charge pulses, SPEED's high-speed analog and digital electronic systems provide information about particle charge, energy and type. The plastic scintillators, the WLS fibers and the electronics are housed in a specially designed aluminum casing that is space-qualified, depicted in Figure 3. Some important specifications of SPEED are detailed in Table 3.

### 4.2. Unique Design Features

SPEED features the following unique characteristics:
- *Large active area*: SPEED features a large active area and geometric factor, which can provide accurate measurement of charged particle flux. This enables us to measure particle burst events of relatively low fluxes

- *High temporal resolution*: SPEED's fast plastic scintillators, PMT's and electronics provide a final data product with a temporal resolution of 0.1 seconds. This can sufficiently characterize the time-characteristics of particle bursts.
- *High energy:* SPEED can measure the energies of protons and electrons over a wide energy range with a high efficiency. The detector can also differentiate between the electrons and protons that lie within the energy range.

Table 3: Major Specifications of SPEED

| Energy range and type of particle | Proton : 17 to 100 MeV |
| --- | --- |
| | Electron : 1 to 15 MeV |
| Energy resolution (of the final data product) | Protons : 5 MeV |
| | Electrons : 1 MeV |
| Particle identification | Categorize incident particles into protons, electrons and 'others' |
| Temporal resolution | Fine: 0.1s; Coarse: 3s |
| Active opening area | 225 x 225 mm$^2$ |
| Field of view | 178.5$^0$ |
| Overall dimensions | 270 X 269 X 88 mm$^3$ |
| Mass (10% Margin) | 7 Kg |
| Peak power consumption | 2.99 W |
| Average power consumption | 1.46 W |
| Shielding method | Active shielding using plastic scintillation detectors |

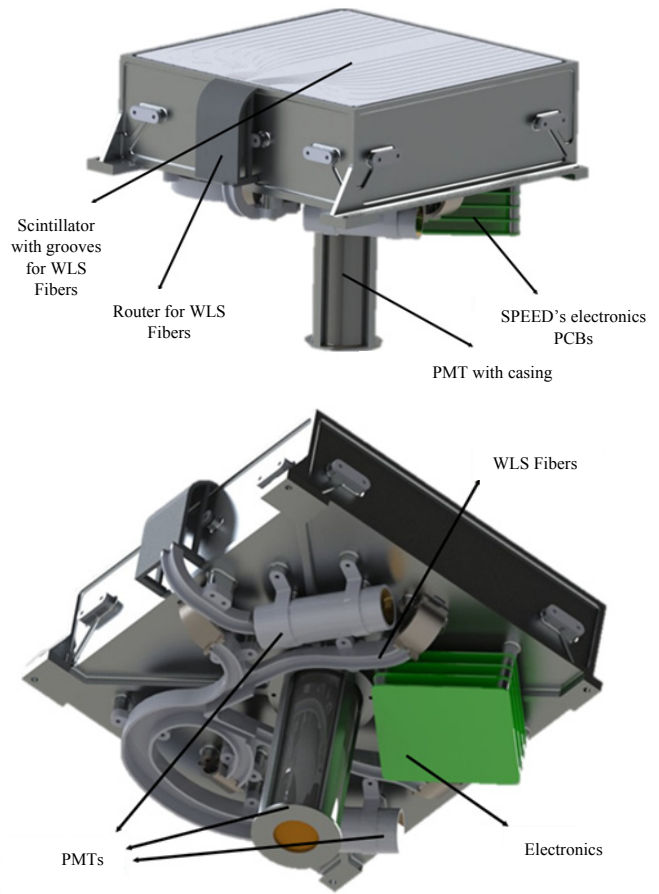

Figure 3: Top view (above) and bottom view (below) of SPEED

## 5. Conclusion

Previous studies by Sgrigna [10] suggest that the particle bursts occur few hours before the earthquake main shock. This could be due to the triggering of particle bursts due to a very high increase in intensity of ULF emission from the seismically active region just a few hours before the earthquake, as observed in Loma Prieta in 1989 [13] and Haiti in 2010 [14].

This short term precursory nature of the phenomenon and the fact that it can give clues to the location of the epicenter of the impending earthquake makes it a potential candidate for short-term earthquake prediction. The observation of particle bursts will have to be done in tandem with measurements of other precursors (such as generation of ULF waves, ionosphere density changes, thermal and stress anomalies on the ground etc.) to improve the confidence level of the prediction.

A few other single satellite missions and satellite constellations are under development around the world to aid earthquake prediction studies, such as China Seismo-Electromagnetic Satellite (CSES) and TwinSat Project. The IITM nano-satellite will be another such mission that will contribute valuable data to this growing field of space research.

**Acknowledgement**

We wish to express our gratitude to Dr. Manju Sudhakar and Dr. Ramakrishna Sharma from ISRO, Prof. Kajari Mazhumdar and Ms. Mandakini Patil from Tata Institute of Fundamental Research, Dr. P. Sreekumar, Dir. of Indian Institute of Astrophysics, and Muriel Richard, Dy. Dir. of Swiss Space Centre, for their support in terms of reviews and technical expertise throughout the mission development phase of the IIT Madras nano-satellite Program. We also acknowledge Prof. P. C. Deshmukh and Prof. S. Santhakumar from IIT Madras, and Dr. Umesh R. Kadhane from Indian Institute of Space Science and Technology for mentoring us from the beginning of the program's development. We acknowledge IITM and IITM Alumni (Batch of 85, SEER Akademi, Signion Systems Pvt. Ltd., and ACSYS Software Pvt. Ltd.) for providing the funding for this program.